\documentclass[conference]{IEEEtran}
\IEEEoverridecommandlockouts
\usepackage{cite}
\usepackage{amsmath,amssymb,amsfonts}
\usepackage{graphicx}
\usepackage{textcomp}
\usepackage{xcolor}
\usepackage{graphicx}
\usepackage{threeparttable}
\usepackage{amsmath}
\usepackage{amsthm}
\usepackage{algorithmicx}
\usepackage{algorithm}
\usepackage{algpseudocode}
\usepackage{booktabs}
\def\BibTeX{{\rm B\kern-.05em{\sc i\kern-.025em b}\kern-.08em
    T\kern-.1667em\lower.7ex\hbox{E}\kern-.125emX}}

\begin{document}
\title{CBlockSim: A Modular High-Performance Blockchain Simulator
\thanks{This work was supported by the China Scholarship Council (File Number: 202006070051).}
\thanks{The short version of this paper has been accepted by 2022 IEEE International Conference on Blockchain and Cryptocurrency as a short paper.}
}
\author{
    \IEEEauthorblockN{Xuyang Ma\IEEEauthorrefmark{1}, Han Wu\IEEEauthorrefmark{2},
    Du Xu\IEEEauthorrefmark{1}, Katinka Wolter\IEEEauthorrefmark{3},\\}
    \IEEEauthorblockA{\IEEEauthorrefmark{1}\textit{School of Information and Communication Engineering, University of Electronic Science and Technology of China}\\
    Chengdu, China\\
    xuyangm7@gmail.com, xudu@uestc.edu.cn}
    \IEEEauthorblockA{\IEEEauthorrefmark{2}\textit{School of Computing, Newcastle University}, Newcastle Upon Tyne, United Kingdom\\
    han.wu@newcastle.ac.uk}
    \IEEEauthorblockA{\IEEEauthorrefmark{3}\textit{Department of Mathematics and Computer Science, Freie Universität Berlin}, Berlin, Germany\\
    katinka.wolter@fu-berlin.de}
}


\maketitle

\begin{abstract}
Blockchain has attracted much attention from both academia and industry since emerging in 2008. Due to the inconvenience of the deployment of large-scale blockchains, blockchain simulators are used to facilitate blockchain design and implementation. We evaluate state-of-the-art simulators applied to both Bitcoin and Ethereum and find that they suffer from low performance and scalability which are significant limitations. To build a more general and faster blockchain simulator, we extend an existing blockchain simulator, i.e. BlockSim. We add a network module integrated with a network topology generation algorithm and a block propagation algorithm to generate a realistic blockchain network and simulate the block propagation efficiently. We design a binary transaction pool structure and migrate BlockSim from Python to C++ so that bitwise operations can be used to accelerate the simulation and reduce memory usage. Moreover, we modularize the simulator based on five primary blockchain processes. Significant blockchain elements including consensus protocols (PoW and PoS), information propagation algorithms (Gossip) and finalization rules (Longest rule and GHOST rule) are implemented in individual modules and can be combined flexibly to simulate different types of blockchains. Experiments demonstrate that the new simulator reduces the simulation time by an order of magnitude and improves scalability, enabling us to simulate more than ten thousand nodes, roughly the size of the Bitcoin and Ethereum networks. Two typical use cases are proposed to investigate network-related issues which are not covered by most other simulators.
\end{abstract}


\section{Introduction}
Bitcoin~\cite{Nakamoto2008} started in 2008 and was the first successful system using a blockchain as core technology. Bitcoin, like several other similar systems, has an underlying peer-to-peer network composed of many connected nodes. Each node holds a copy of the ledger, which is an ordered sequence of blocks. The nodes periodically reach consensus on the next valid block using a consensus protocol. Hash pointers connecting the blocks guarantee immutability of the block contents, i.e. the transactions. While transactions in Bitcoin are rather simple, Ethereum~\cite{Ethereum2013} supports smart contracts, Turing complete programs, as transactions.  

With many attractive features including tamper resistance, high security, decentralization and smart contracts, blockchain systems have received a lot of attention by academia and industry. We found that blockchain technology has been widely used in many different fields such as finance~\cite{Schar2021,Kowalski2021}, internet of things (IoT)~\cite{Yang2021,Li2021}, the energy market~\cite{JYang2021,Foti2021}, and healthcare~\cite{Miyachi2021,Zou2021}. While researchers have a huge interest in blockchain, popular blockchains like Bitcoin and Ethereum do not support flexible modification and configuration. Therefore, many blockchain simulators, from early Bitcoin Simulator~\cite{Gervais2016} to recent BlockSim~\cite{Alharby2019} and SimBlock~\cite{Aoki2019}, have been proposed to simulate blockchain. 

We have evaluated state-of-the-art blockchain simulators and noticed their limited performance and scalability. Some simulators such as Bitcoin Simulator and SimBlock do not explicitly include transactions in order to accelerate the simulation. But some research requires simulation for transactions to estimate the run time of smart contract~\cite{Putz2021} or the throughput of blockchain~\cite{Yu2021}. Other simulators such as BlockSim:Faria~\cite{Faria2019} and BlockSim:Alharby~\cite{Alharby2019} aim at including much more detail but suffer from high run time and low scalability. Therefore recent research experimented in different ways rather than using existing blockchain simulators. Some work~\cite{Putz2021, Yu2021} deployed a small-scale blockchain on a few computers or servers which is very different from the real blockchain environment. Other work~\cite{Kumar2021,Mishra2021} ran experiments on Ethereum testnets (Rinkeby or Ropsten) where the configuration is unknown and blockchain settings can not be adjusted. The work in~\cite{Chang2021,Piao2021} did not apply experiments in a real blockchain but instead used theoretical analysis. 

\textit{Contributions:} To maintain a detailed simulation and achieve high performance and scalability, we extend BlockSim:Alharby to be a flexible high-performance blockchain simulator named CBlockSim. Our contributions are as follows.
\begin{enumerate}
    \item We add a network module integrated with a network generation algorithm and a Dijkstra-based algorithm to generate a realistic blockchain network and simulate information propagation efficiently.
    \item We rebuild BlockSim:Alharby in C++ and design a binary transaction pool data structure so that we can adopt bitwise operations in C++ to accelerate the simulation. The experiments compare CBlockSim with other state-of-the-art simulators, indicating that CBlockSim shortens the run time by an order of magnitude and increases scalability by an order of magnitude.
    \item We modularize the simulator based on five blockchain processes. The simulator can be configured to represent different blockchains, i.e. different consensus protocols, different information propagation algorithms and finalization rules implemented in individual modules.
    \item We propose two typical use cases towards network-related issues that most other simulators do not cover due to the lack of precise network simulation.
\end{enumerate}

\section{Related Work}
Since Bitcoin and Ethereum are the two most popular blockchains, most blockchain simulators aim to simulate either or both of them. Bitcoin Simulator~\cite{Gervais2016} built in C++ and ns3 are the most widely used simulators for Bitcoin. VIBES~\cite{Stoykov2017} is another large-scale Bitcoin simulator built in Scala. An Ethereum simulator eVIBES~\cite{Deshpande2018} was proposed inspired by VIBES. SimBlock~\cite{Aoki2019} is a simulator for Bitcoin built in Java, supporting compact block relay. It is also expected to support simulation for Ethereum in the future. BlockSim~\cite{Faria2019} proposed by Faria and Correia aims to simulate both Bitcoin and Ethereum with high flexibility. Another simulator also named BlockSim~\cite{Alharby2019} proposed by Alharby and van~Moorsel is built in Python and can simulate both blockchains as well.

All these simulators are analyzed in \cite{Paul2021}. Some simulators cannot simulate several of the blockchain components. VIBES, eVIBES and BlockSim:Alharby adopt a fixed information propagation delay so that they can not simulate realistic networks. Bitcoin Simulator and SimBlock ignore simulation of transactions but concentrate on blocks. Some simulators have limited performance. BlockSim: Faria is only used to simulate a small-scale blockchain. BlockSim:Alharby suffers from high run time when simulating transactions. Our proposed CBlockSim aims at performance and scalability and simulates a significant number of blockchain components.

\section{CBlockSim Architecture}
\begin{figure}[!hbt]
	\centering
		\includegraphics[width=3.5in]{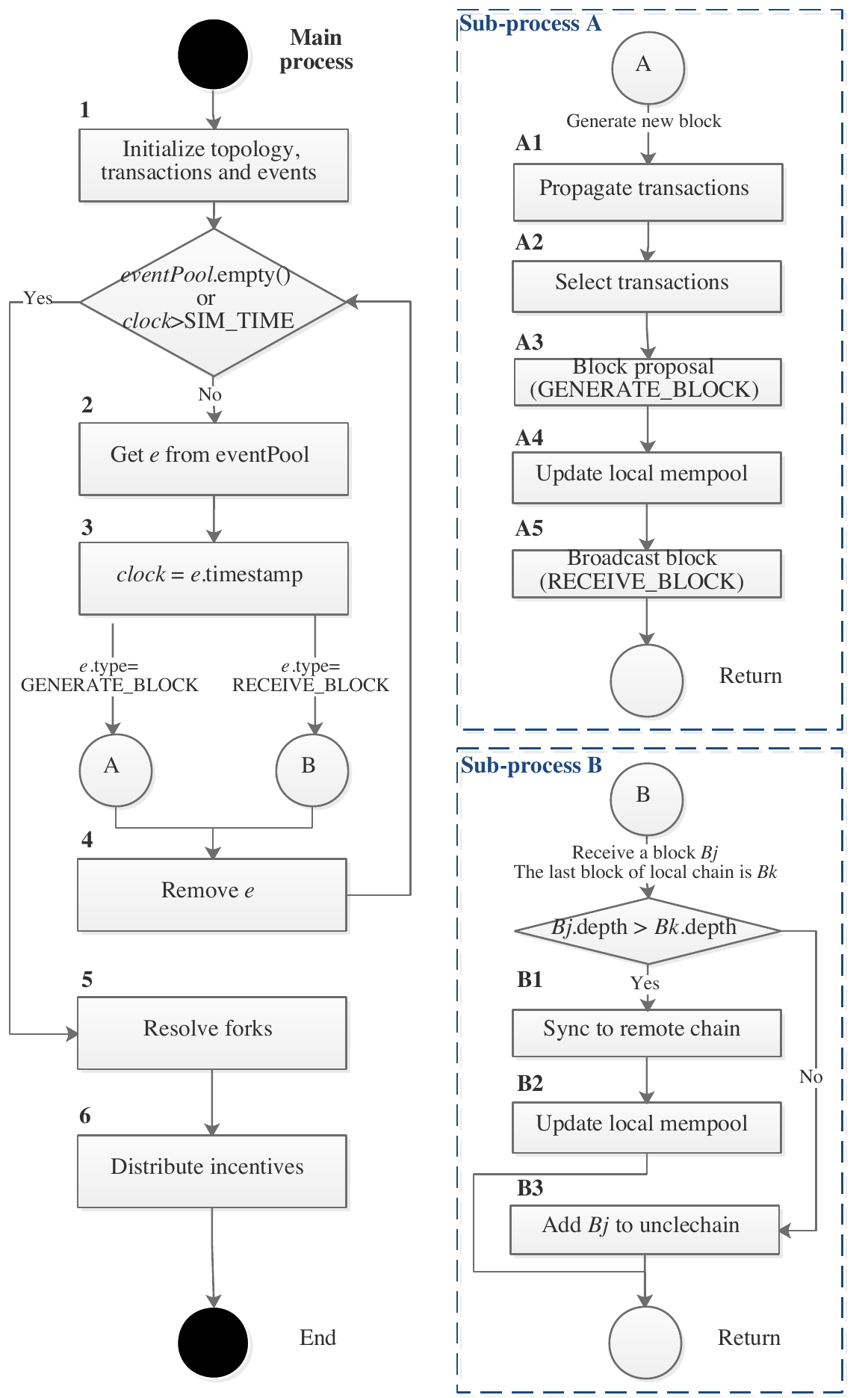}
	\caption{The flowchart of CBlockSim}
	\label{FIG:flowchart}
\end{figure}

As illustrated in Figure~\ref{FIG:flowchart}, CBlockSim is built on a discrete-event simulation model. A queue $eventPool$ is adopted to manage two types of events, GENERATE\_BLOCK event and RECEIVE\_BLOCK event. The global timestamp $clock$ which is no more than SIM\_TIME (the simulation time set by the user) updates with the change of the timestamp of each event.  Process A deals with GENERATE\_BLOCK event and process B deals with RECEIVE\_BLOCK event. We modularize CBlockSim based on the following five  blockchain processes~\cite{Xiao2020}.

\subsection{Block Proposal}
Block Proposal (step A3 in Figure~\ref{FIG:flowchart}) is responsible for block generation according to different consensus protocols. It is an individual module for simulating the generation time of every generated block. We implement an algorithm to simulate two mainstream consensus protocols, PoW and PoS. The PoW simulation algorithm is consistent with the algorithm in BlockSim:Alharby. Suppose there are $N$ miners where \textbf{N}=\{\(1, 2, \cdots, N\)\}. The hash power of a miner $i$ is denoted by $H_i$. The fraction of blocks generated by miner $i$ should be proportional to \(\frac{H_i}{\sum_{j \in \textbf{N}}H_j}\). Therefore the exponential distribution is adopted to simulate the time between blocks as shown in PoW function in Algorithm \ref{BlockProposal}. 

PoW is the most common consensus protocol of many blockchains but it is often accused of consuming too much energy. PoS is a promising energy-saving alternative consensus protocol. Miners compete with each other based on their holding stakes instead of computing power. We simulate the PoS protocol as in PeerCoin~\cite{Zhao2021} which considers both the stake owned by the miner and the duration that the miner has held on to it. Suppose the stake owned by miner $i$ is $S_i$ and the duration that the miner has held the stake is denoted by $Age_i$. Then the fraction of blocks that miner $i$ generates should be proportional to \(\frac{S_i \cdot Age_i}{\sum_{j \in \textbf{N}}S_j \cdot Age_j}\). The PoS function in Algorithm \ref{BlockProposal} presents the simulation of this process.

\begin{algorithm}[!hbt]
\caption{Block Proposal}
\label{BlockProposal}
\begin{algorithmic}[1]
  \Statex $BlockInterval$: the time interval between two blocks;
  \Function{PoW}{$H_i$, $currentTimestamp$}
  \State \(\lambda \leftarrow \frac{H_i}{\sum_{j \in \textbf{N}}H_j} \cdot \frac{1}{BlockInterval}\);
  \State The time took to mine a block: \(T \sim Exp(\lambda)\);
  \State The generation time of the next block: \(genTimestamp \leftarrow currentTimestamp + T\);
  \State \Return \(genTimestamp\);
  \EndFunction
  \Statex
  \Function{PoS}{$S_i$, $Age_i$, $currentTimestamp$}
  \State \(\lambda \leftarrow \frac{S_i \cdot Age_i}{\sum_{j \in \textbf{N}}S_j \cdot Age_j} \cdot \frac{1}{BlockInterval}\);
  \State The time took to mine a block: \(T \sim Exp(\lambda)\);
  \State The generation time of the next block: \(genTimestamp \leftarrow currentTimestamp + T\);
  \State \Return \(genTimestamp\);
  \EndFunction
\end{algorithmic}
\end{algorithm}

\subsection{Information Propagation}

\begin{figure*}
	\centering
		\includegraphics{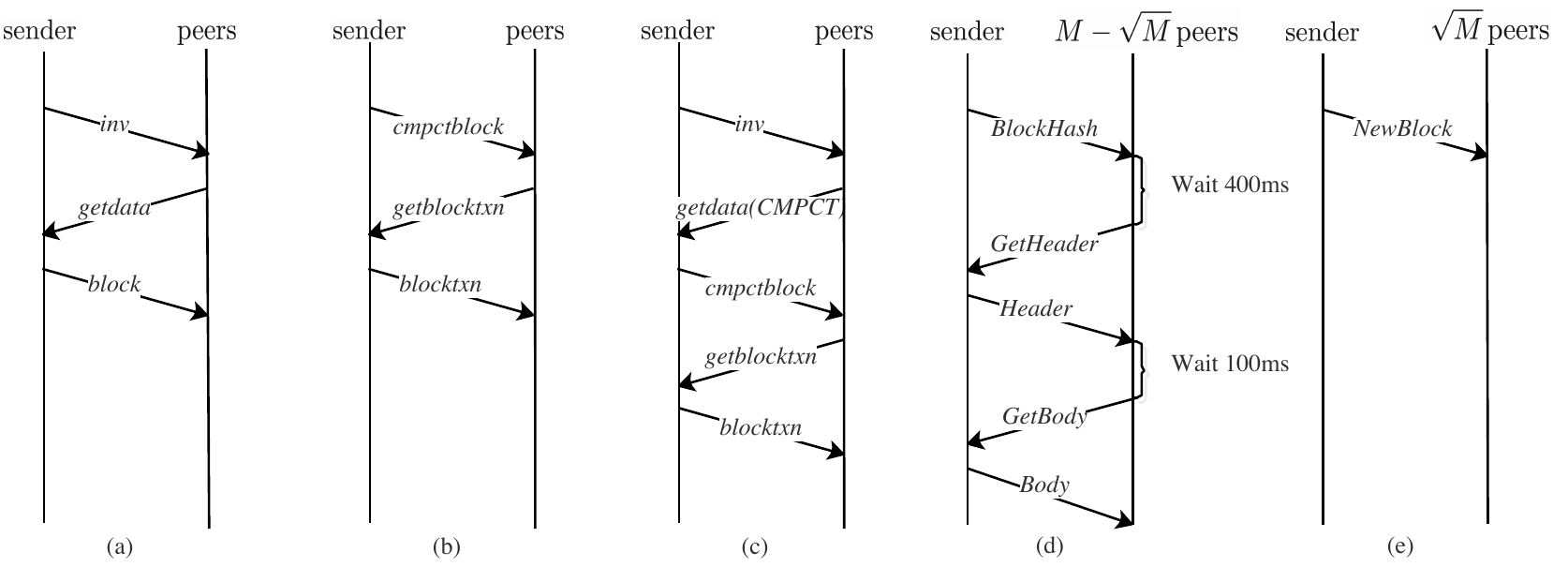}
	\caption{(a)(b)(c) present the block transmission in Bitcoin; (d)(e) present the block transmission in Ethereum}
	\label{FIG:blockchainTransmission}
\end{figure*}

While the underlying network is a key part of a blockchain and could seriously affect the performance of the blockchain~\cite{Hao2019, Rohrer2019}, in many simulators there is no realistic network model. BlockSim:Alharby adopted the exponential distribution to simulate the block propagation delay. 
Instead, we add a network module to simulate information propagation in detail (step 2 in Figure~\ref{FIG:flowchart}). It first generates a realistic blockchain network topology and then calculates the block propagation delay.

The Bitcoin network is an unstructured peer-to-peer network which can be simulated using a random graph model~\cite{Shahsavari2020}. The Ethereum network is a structured peer-to-peer network and has been proven to have the small-world property~\cite{Gao2019,WangT2021,Maeng2021}. It can be simulated using the Watts-Strogatz model~\cite{Watts1998}. We adopt the method proposed in \cite{Song2014} to generate a random network and a small-world network. The average degree $d$ and a topology controlling parameter \(\beta\) are used to control the generation of different networks. We set $d$ to 12~\cite{Miller2015} and \(\beta\) to 1 to generate a random graph (Bitcoin), and set $d$ to 19.747 and \(\beta\) to 0.24 to generate a small-world network (Ethereum) as observed in \cite{Maeng2021}. As presented in Algorithm~\ref{NetworkGeneration}, the $GetProbability$ function calculates the probability of a connection between node $i$ and node $j$. The $GenerateNetwork$ function compares the generated random number and the probability to determine whether two nodes are connected or not.

\begin{algorithm}[!hbt]
\caption{Network Generation}
\label{NetworkGeneration}
\begin{algorithmic}[1]
  \Statex $N$: the number of nodes;
  \Statex $\beta$: topology controlling parameter;
  \Statex $d$: average number of connections;
  \Statex $G$: graph containing all the nodes;
  \Function{GetProbability}{$N$, $\beta$, $d$, $i$, $j$}
    \State \(dis \leftarrow |i-j|\);
    \State \(edgeDensity \leftarrow \frac{d}{N-1}\);\algorithmiccomment{$p_0$ in \cite{Song2014}}
    \State \(maxDistance \leftarrow \lfloor \frac{N}{2} \rfloor\);\algorithmiccomment{$D_{max}$ in \cite{Song2014}}
    \If{\(edgeDensity-\frac{min(dis, N-dis)}{maxDistance} \geq 0\)}
        \State \Return \(\beta \cdot (edgeDensity-1)+1\);
    \Else
        \State \Return \(\beta \cdot edgeDensity\);
    \EndIf
  \EndFunction
  \Statex
  \Function{GenerateNetwork}{$N$, $\beta$, $d$}
    \State Create a graph $G$ with $N$ nodes and no edge;
    \For{\(i, j \in \textbf{N}\)}
        \State \(p \leftarrow GetProbability(N, \beta, d, i, j)\);
        \State \(r \leftarrow random(0, 1)\);
        \If{\(i = j\) or \(r > p\)}
            \State \(G.edge(i, j) \leftarrow 0\);
        \Else
            \State \(G.edge(i, j) \leftarrow 1\);
        \EndIf
    \EndFor
    \State \Return $G$;
  \EndFunction
\end{algorithmic}
\end{algorithm}

\begin{algorithm}[!hbt]
\caption{Information Propagation}
\label{InformationPropagation}
\begin{algorithmic}[1]
  \Statex \textbf{Input:}
  \Statex $sender$: the sender of the transaction or block;
  \Statex $G$: graph containing all the nodes;
  \Statex \textbf{Output:}
  \Statex $D$: transmission delays from $sender$ to other nodes;
  \State Set priority queue \(Q \leftarrow \varnothing\), updated node vector \(V \leftarrow \varnothing\);
  \For {\(v \in G\)}
    \State \(D[v] \leftarrow \infty\);
    \If {$v$ is a neighbor of $sender$}
        \State \(Q.add(v)\);
        \State Calculate \(D[v]\) according to Equation~\ref{EQ:BitcoinTransmissionDelay} or \ref{EQ:EthereumTransmissionDelay};
    \EndIf
  \EndFor
  \State \(D[sender] \leftarrow 0\);
  \State \(V.add(sender)\);
  \While {$Q$ is not empty}
    \State \(u \leftarrow Q.pop()\);
    \State \(Q.remove(u)\);
    \For {each neighbor $v$ of $u$}
        \State Calculate \(d_{uv}\) (delay between $u$ and $v$) according to Equation~\ref{EQ:BitcoinTransmissionDelay} or \ref{EQ:EthereumTransmissionDelay};
        \State \(D[v] \leftarrow min(D[v], d_{uv})\);
    \EndFor
  \EndWhile
\end{algorithmic}
\end{algorithm}

We propose an algorithm based on the Dijkstra algorithm to calculate the block propagation delay as shown in Algorithm~\ref{InformationPropagation}. Step A5 in Figure~\ref{FIG:flowchart} broadcasts the generated block according to the calculated block propagation delay. The difference between the calculation of the delay of Bitcoin and Ethereum is due to the difference of transmission protocols. (a)(b)(c) in Figure~\ref{FIG:blockchainTransmission} illustrate three block transmission modes in the Bitcoin network, where (a) is the legacy relaying mode. The node sends the $inv$ message to notify all its peers of the arrival of a new block. The peer replies with the $getdata$ message to request the new block. (b) and (c) are the high bandwidth relaying and low bandwidth relaying modes proposed in Compact Block Relay (CBR)~\cite{BIP0152}. 

In the high bandwidth relaying mode the node sends $cmpctblock$ message to notify the peer of the arrival of a new block. If the peer has all the transactions contained in the new block, it will reconstruct the block locally and no further transmission is needed. But if some transactions are not available, a $getblocktxn/blocktxn$ roundtrip is needed to obtain remaining transactions. 

In the low bandwidth relaying mode a peer receives an $inv$ message and returns a $getdata(CMPCT)$ message to request the header and short transaction IDs. As in the high bandwidth relaying mode, the peer will try to reconstruct the block locally unless necessary transactions are not available. Thus, in the best case only one $cmpctblock$ message is sent so that one latency is needed. In the worst case five messages are sent and the latency is counted five times. In our simulator we count the latency three times which is the average case and ignore the message size because in most cases small-size messages instead of the entire block are sent to request missing transactions. So the transmission delay between two adjacent nodes can be denoted by Equation~\ref{EQ:BitcoinTransmissionDelay}. $D$ indicates the transmission delay. $L$ indicates the latency. $size$ indicates the block size and $pd$ indicates the block processing delay for 1 Mb block.

\begin{equation}
    D = 3 \cdot L + pd \cdot size.
    \label{EQ:BitcoinTransmissionDelay}
\end{equation}

(d) and (e) illustrate the transmission in Ethereum network. A node sends the complete block only to a small fraction of its connected peers (usually the square root of the total number of peers) and sends the hash of the new block to other peers. A peer which receives the hash at first waits for $400$~ms. Then it sends the $GetHeader$ message to request the block header. After receiving the block header, it waits for $100$ ms and sends the $GetBody$ message to request the block body. The sender and its peer send messages five times in total. So the transmission delay can be described by Equation~\ref{EQ:EthereumTransmissionDelay}. $B$ denotes the bandwidth and $M$ denotes the number of connected peers. We use the $random$ function to simulate the random choice of peers of an Ethereum node.

\begin{equation}
    D = \begin{cases}L + size\cdot (B^{-1} + pd), & random(1, M) \leq \sqrt{M} 
    \\ 5 \cdot L + size\cdot (B^{-1} + pd), & random(1, M) > \sqrt{M}
    \end{cases}
    \label{EQ:EthereumTransmissionDelay}
\end{equation}

\subsection{Block Validation}
When receiving a new block from other nodes, the receiver needs to check the generation proof and validate and execute all the included transactions, causing a considerable block processing delay which affects the block propagation delay (Equation \ref{EQ:BitcoinTransmissionDelay} and \ref{EQ:EthereumTransmissionDelay}). We adopt the linear model in \cite{Shahsavari2020} to estimate this delay. Compared to the constant or random variable used in other simulators, this active model can reflect the impact of block size on the block propagation delay better.

BlockSim:Faria used $229$ ms and $240$ ms as the average overall block processing delays of Ethereum and Bitcoin based on measurements. We take the two values to estimate block processing delays on each node. First, Considering the generated Bitcoin topology in Section III(B) of which the average path length is about 4 and the Bitcoin block size is 1.22 MB, the block processing delay for 0.1 MB Bitcoin block will be \(240 / 4 / 1.22 / 10 \approx 5\) ms, which is close to the theoretical estimated block processing delay (4 ms) in \cite{Shahsavari2020}. Then using the same way to infer the block processing delay of Ethereum, the average path length of the generated Ethereum topology is 3.8 which is close to the estimated value (3.7) in \cite{WangT2021}. The average Ethereum block size in 2020 is about 0.0225 MB. So the estimated processing delay for 0.1 MB Ethereum block is \(229 / 3.8 / 0.0225 / 10 \approx 268\) ms.

\subsection{Block Finalization}
Block finalization reaches the agreement on the acceptance of blocks. Most blockchains adopt the longest chain rule to determine the accepted chain but there are still some other choices such as GHOST which is used in the early version of Ethereum. While BlockSim:Alharby supports the longest chain rule, we include the GHOST rule in our simulator as well. Process~B in Figure~\ref{FIG:flowchart} illustrates the process of the longest chain rule.

\subsection{Incentive Mechanism}
The incentive mechanism decides the way to distribute block rewards and transaction fees. Bitcoin only rewards the miners who generate blocks in the longest chain. Ethereum rewards the miners who generate blocks in the accepted chain or uncle blocks. Step~9 in Figure~\ref{FIG:flowchart} distributes rewards according to different incentive mechanisms.

\section{Transaction Pool Design}

Besides the above block-related processes, there are some significant transaction-related steps that seriously affect the performance of the simulator. First, the number of transactions is much larger than the number of blocks. Further, the execution times of transaction-related operations such as steps B2 and B5 in Figure~\ref{FIG:flowchart} will increase as the number of nodes increases. Suppose there are $N$ nodes in total, every generated new block will be propagated to all other nodes according to the Algorithm~\ref{InformationPropagation} so each GENERATE\_BLOCK event will trigger $N$-1 RECEIVE\_BLOCK events. Then the frequency of executing step B2 or B5 will increase with the growth of the number of nodes. Therefore, we make a special design on the transaction pool structure to avoid massive iterative operations, using bitwise operation in C++.

\begin{figure}[!hbt]
	\centering
		\includegraphics[width=0.8\linewidth]{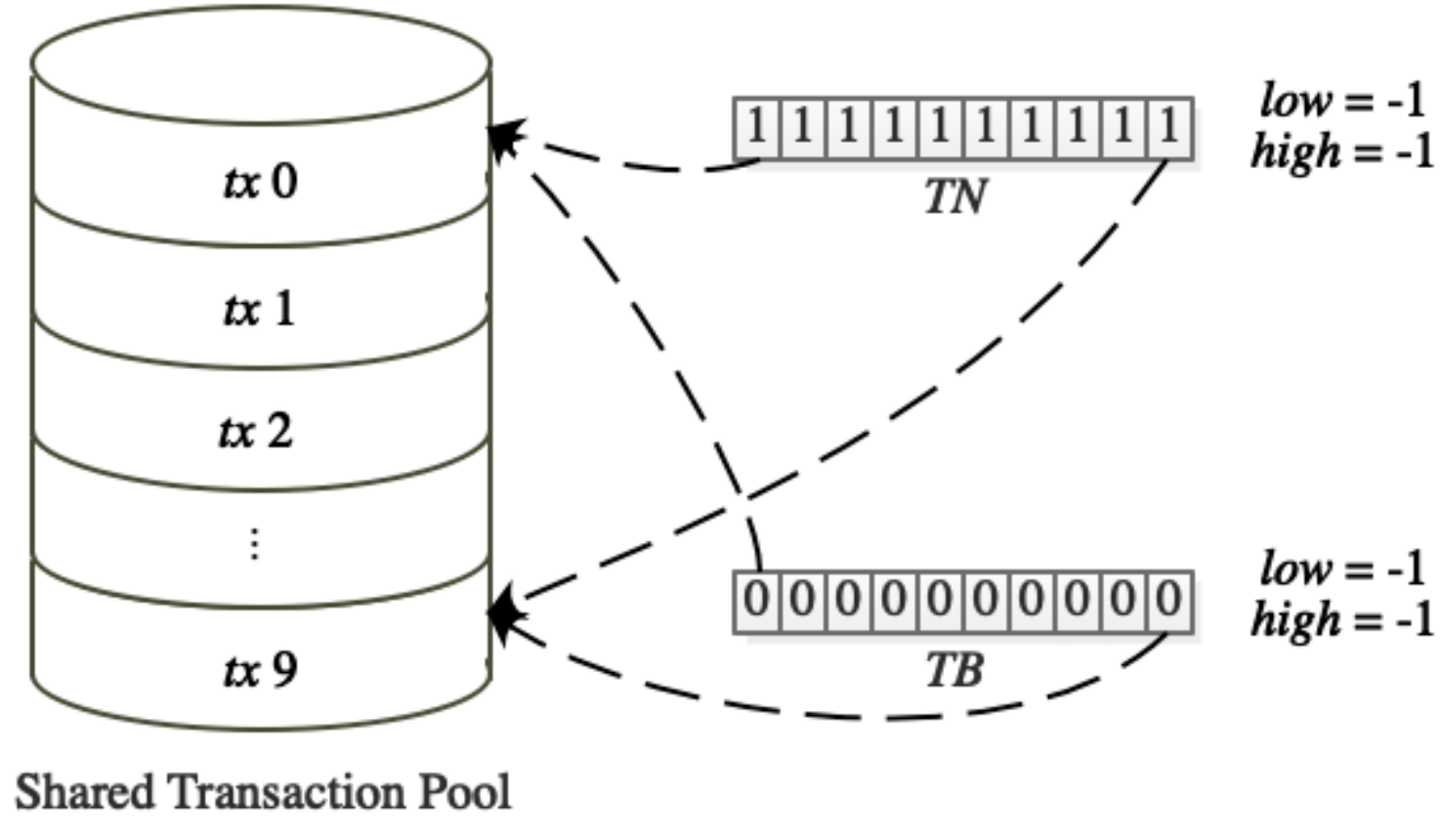}
	\caption{Transaction storage structure}
	\label{FIG:storage}
\end{figure}

As illustrated in Figure~\ref{FIG:storage}, we first store the complete transactions in a transaction pool which is shared by all the nodes. These complete transactions will be stored in memory only once and will be sorted by the generation timestamp. Every node has a {\em mempool} to record received transactions which are not included in the local chain. We use the binary sequences, \(TN\) (transactions in the mempool of the node) and \(TB\) (transactions in the block), to represent transactions in each node and block. Two pointers, $low$ and $high$, are used to determine the valid range of transactions. The length of the two binary sequences is equal to the total number of transactions generated during simulation. Only the transactions between $low$ and $high$ are available. 

For example, if ten transactions are generated during simulation, \(TN\) and \(TB\) will have ten bits. The leftmost digit represents the first generated transaction and the rightmost digit represents the last generated transaction. If a digit is set to 1 and locates between $low$ and $high$, it means the block or the mempool of the node contains this transaction. At the beginning, $low$ and $high$ in every node and block are set to -1 which means no transactions are included in any mempool or block. All the bits of $TN$ are initialized to 1 while all the bits of $TB$ are initialized to 0. 

\begin{figure}[!hbt]
	\centering
		\includegraphics[width=0.9\linewidth]{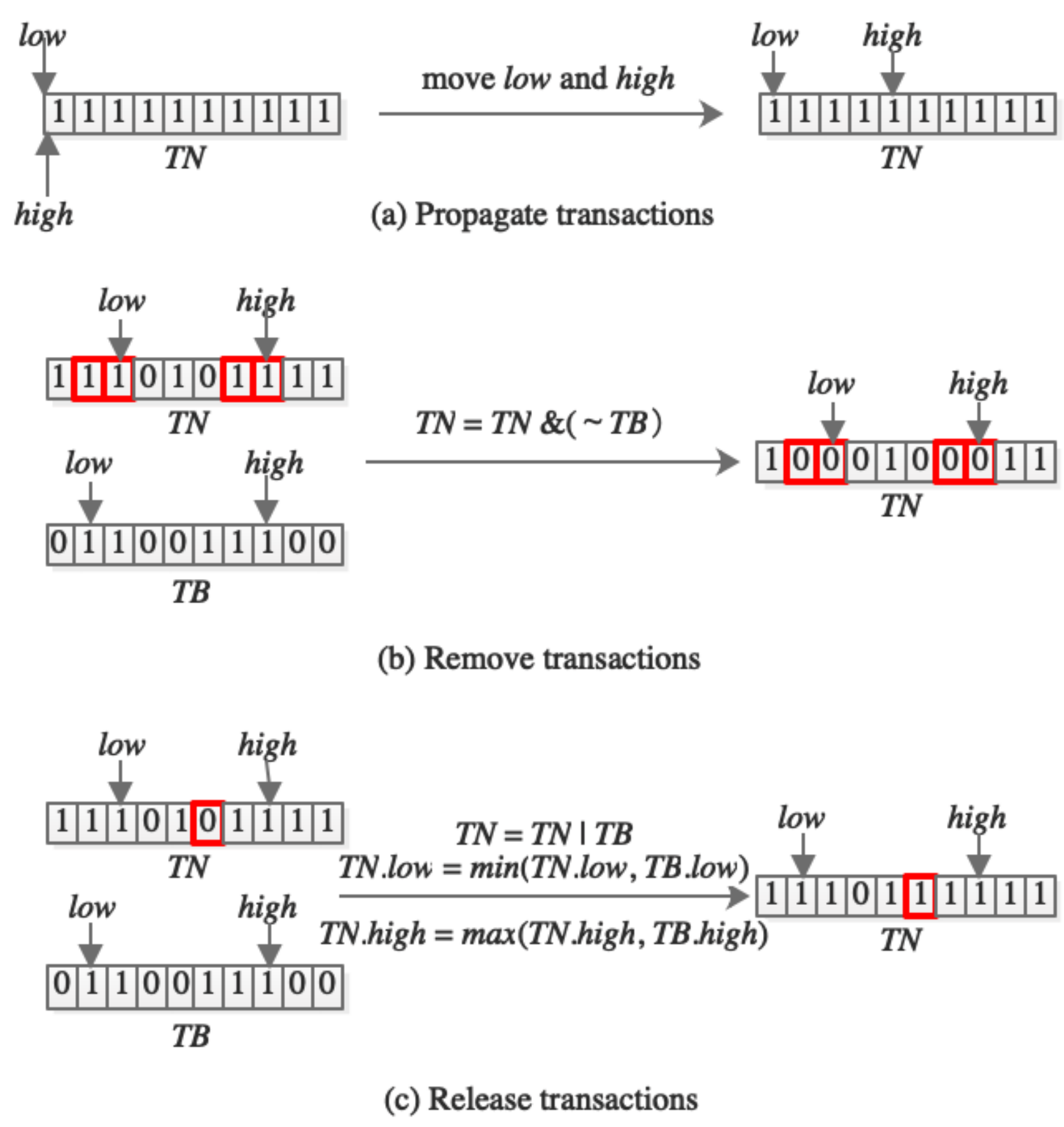}
	\caption{(a) Move $low$ and $high$ to propagate transactions (b) Remove transactions included in a block from the mempool (c) Release transactions back to the mempool}
	\label{FIG:txn}
\end{figure}

Suppose the number of nodes is $N$ and the number of transactions is $M$. Three following operations benefit from the transaction pool structure design:

\begin{enumerate}
    \item Transaction propagation (step A1 in Figure~\ref{FIG:flowchart}): as illustrated in Figure~\ref{FIG:txn}(a), before generating a new block, transactions that were generated earlier than the block generation timestamp should be propagated to all nodes. We only need to move $low$ and $high$ to include transactions that were generated before current timestamp.
    \item Transaction removal (step A4 and B2 in Figure~\ref{FIG:flowchart}): when a block is included in the local chain of a node, the node needs to remove the transactions that are included in this block from the mempool. Figure~\ref{FIG:txn}(b) illustrates this process which is a bitwise NAND operation.
    \item Transaction release (step B5 in Figure~\ref{FIG:flowchart}): when a node receives a block from another node with a longer chain, the receiver needs to synchronize its local chain with the longer chain. Some blocks in the local chain will become stale or uncle blocks. The transactions included in these blocks should be released back to the mempool to be included again later. Figure~\ref{FIG:txn}(c) illustrates this process which is a bitwise OR operation.
\end{enumerate}

\section{CBlockSim Validation}

We validate CBlockSim by comparing the simulation results generated by CBlockSim with the observed realistic data of Bitcoin and Ethereum. To deal with different application scenarios, CBlockSim can be configured to enable or disable network simulation. Users who are not willing to simulate network details can use an exponential distribution to simulate a delay as in BlockSim:Alharby.

We first simulate Bitcoin using data from 2020 and compare the result with the realistic data presented in \cite{Paul2021}. Then we simulate Ethereum observed in \cite{Maeng2021} and compare the result with the data obtained from Etherscan\footnote{https://etherscan.io/charts} and Ethstats\footnote{https://ethstats.net/}. The experiment is performed ten times and we take the average value as the result. The simulation configuration is presented in Table~\ref{TB: Configuration}.

\begin{table}[!hbt]
\begin{center}
\caption{Simulation configuration}
\label{TB: Configuration}
\def\arraystretch{1.2}
\begin{threeparttable}
\begin{tabular}{*3{c}}
\toprule
 & \textbf{Bitcoin} & \textbf{Ethereum} \\
\hline
\#Nodes & 11,000 & 8,223 \\
Simulation Time (s) & 600,000 & 86,400 \\
Block Interval (s) & 600 & 13.05 \\
Geographical Node Distribution & \cite{Paul2021} & \cite{Maeng2021} \\
Average Degree & 12 & 19.747 \\
Topology Controlling Parameter & 1 & 0.24 \\
Bandwidth/Latency Distribution & \cite{Aoki2019} & \cite{Aoki2019} \\
Hash Power Distribution & \cite{Nagayama2020} & \cite{Nagayama2020} \\
\bottomrule
\end{tabular}
\end{threeparttable}
\end{center}
\end{table}

Table~\ref{TB: Validation Comparison} compares real data and the simulation output including the average block size, the block propagation delay and the stale/uncle rate. When disabling the network module, the fixed average block propagation delay and block size are used as the input parameter and the result is consistent with the simulated result of BlockSim:Alharby. We set the average block propagation delay to $0.7$ seconds for both Bitcoin and Ethereum. In this condition the simulated 50th percentile of the block propagation delay is approximately $0.5$ seconds which is consistent with real data.

\begin{table}[!hbt]
\caption{Comparison of average block size $s_B$, 50th and 90th percentile of block propagation delay $t_{BPD}$, and stale/uncle rate $r$}
\label{TB: Validation Comparison}
\def\arraystretch{1.2}
\begin{threeparttable}
\begin{tabular}{*5{c}}
\toprule
 & $s_{B}$(MB) & $t_{BPD}^{50th}$ (s)& $t_{BPD}^{90th}$ (s) & $r$ \\
\hline
Bitcoin\tnote{m} & 1.22 & 0.5 & 3.3 & 0.06\%\\
Bitcoin\tnote{a} & 1.21$\pm$0.02 & 0.49$\pm$0.001 & 1.61$\pm$0.003 & 0.11\%$\pm$0.006\%\\
Bitcoin\tnote{b} & 1.22$\pm$0.01 & 0.35$\pm$0.005 & 1.34$\pm$0.002 & 0.08\%$\pm$0.001\%\\
Ethereum\tnote{m} & 0.023 & 0.5 & 1.75 & 5.36\%\\
Ethereum\tnote{a} & 0.024$\pm$0.001 & 0.49$\pm$0.001 & 1.61$\pm$0.002 & 4.80\%$\pm$0.15\%\\
Ethereum\tnote{b} & 0.024$\pm$0.002 & 0.51$\pm$0.002 & 1.72$\pm$0.007 & 5.15\%$\pm$0.16\%\\
\bottomrule
\end{tabular}
\begin{tablenotes}
\footnotesize
\item[m] Observed data.
\item[a] Disable network topology simulation.
\item[b] Enable network topology simulation.
\end{tablenotes}
\end{threeparttable}
\end{table}

The simulator output is close to the real observed data.  When enabling the network topology simulation, either the simulated block propagation delay of Ethereum or Bitcoin is close to the observed one which means the network models we used are relatively precise. 

\section{Performance Evaluation}

In order to evaluate the performance of the proposed simulator, we compare the run time and memory usage of CBlockSim, BlockSim:Faria and BlockSim:Alharby while varying the number of nodes and simulation time. All three simulators perform both Bitcoin and Ethereum simulation and have complete simulation functions. Other popular simulators such as Bitcoin Simulator and SimBlock can only simulate Bitcoin and are short of simulation for transactions. All the experiments are performed to simulate Ethereum as observed in \cite{Maeng2021} on a Linux virtual machine with 1 CPU and 32 GB memory. We use the Ethereum configuration in Table~\ref{TB: Configuration}.

\begin{figure*}[!t]
	\centering
		\includegraphics[width=0.7\linewidth]{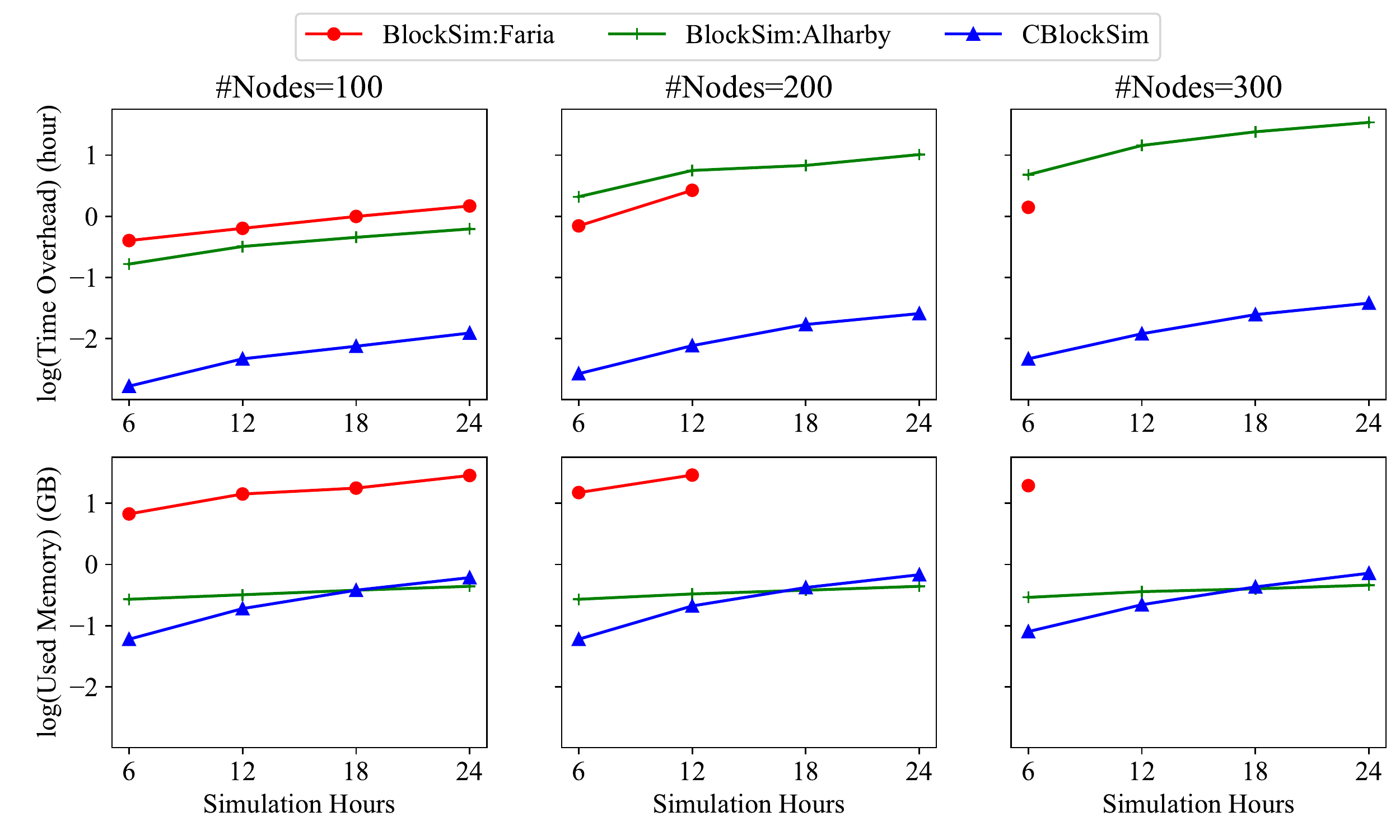}
	\caption{Comparison of time overhead and memory usage}
	\label{FIG:Performance Comparison}
\end{figure*}

We first set the number of nodes to $100, 200$ and $300,$ respectively. The simulation time is set to $6, 12, 18$ and $24$ hours, respectively. Figure~\ref{FIG:Performance Comparison} presents the results. BlockSim:Faria can only simulate a small scale blockchain with short simulation time. The memory usage of BlockSim:Faria is much higher than of the other two simulators. For $200$ nodes and a simulation time of more than $12$ or $300$ nodes and a simulation time of more than $6$, it will run out of memory (more than 32 GB). Both BlockSim:Alharby and CBlockSim can simulate a blockchain well when the number of nodes varies from $100$ to $300$ and the simulation time varies from $6$ to $24$ hours. The difference in memory usage is significant because BlockSim:Faria simulates many details of a blockchain while the other two simulators simplify some components. BlockSim:Alharby shares transactions among all the nodes and uses the exponential distribution to simulate latency. CBlockSim stores complete transactions only once and adopts the binary encoding to represent transactions in each node and block. It further combines a realistic network topology and the shortest path algorithm to simulate network communication.

\begin{figure}[!hbt]
	\centering
		\includegraphics[width=3.5in]{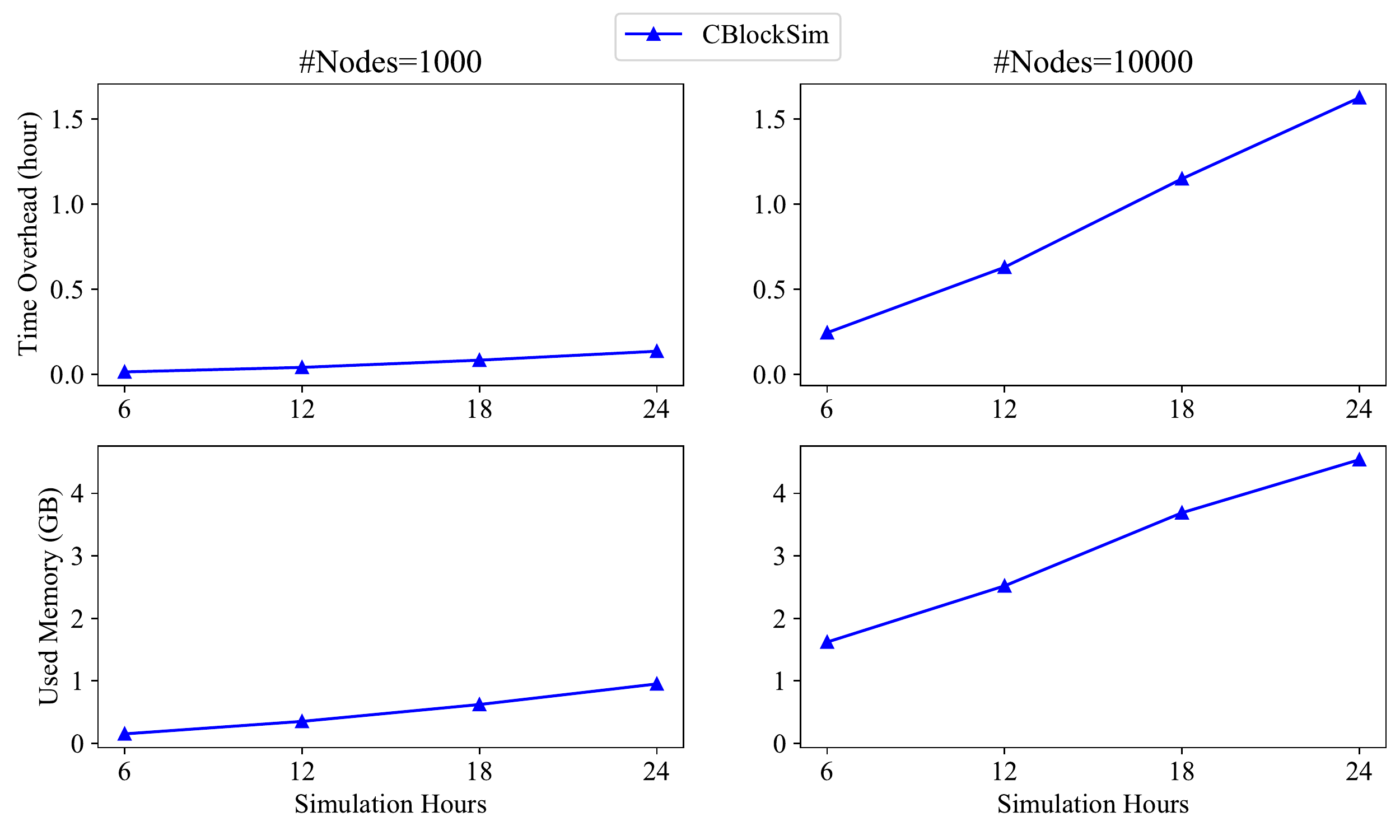}
	\caption{Evaluation of CBlockSim's scalability}
	\label{FIG:Performance Evaluation}
\end{figure}

With increasing number of nodes, a fast growth of the time overhead of BlockSim:Alharby is observed from minutes to hours. In opposition, the time overhead of CBlockSim is low and it increases much slower, from seconds to a few minutes. In Figure~\ref{FIG:Performance Evaluation} we evaluate CBlockSim individually when setting the number of nodes to $1,000$ and $10,000$ because the time overhead of BlockSim:Alharby is too large in this condition. The high time overhead impairs the application of BlockSim:Alharby. The scale of $10,000$ nodes is close to the real blockchain, either Bitcoin or Ethereum. When simulating a blockchain with $10,000$ nodes for $24$ hours, CBlockSim takes about one and half hours and 4.5 GB of memory. This demonstrates that CBlockSim has low time overhead and can be applied to a large-scale blockchain with ten thousand nodes.

\section{Use Cases}

Using CBlockSim we can investigate some network-related issues that many other simulators do not cover. In this section we propose two use cases, investigating the effect of different transmission protocols and the impact of different average node degree on the blockchain  network.

\subsection{Compact Block Relay v.s. Eth Wire Protocol}
\begin{figure}[!hbt]
	\centering
		\includegraphics[width=0.8\linewidth]{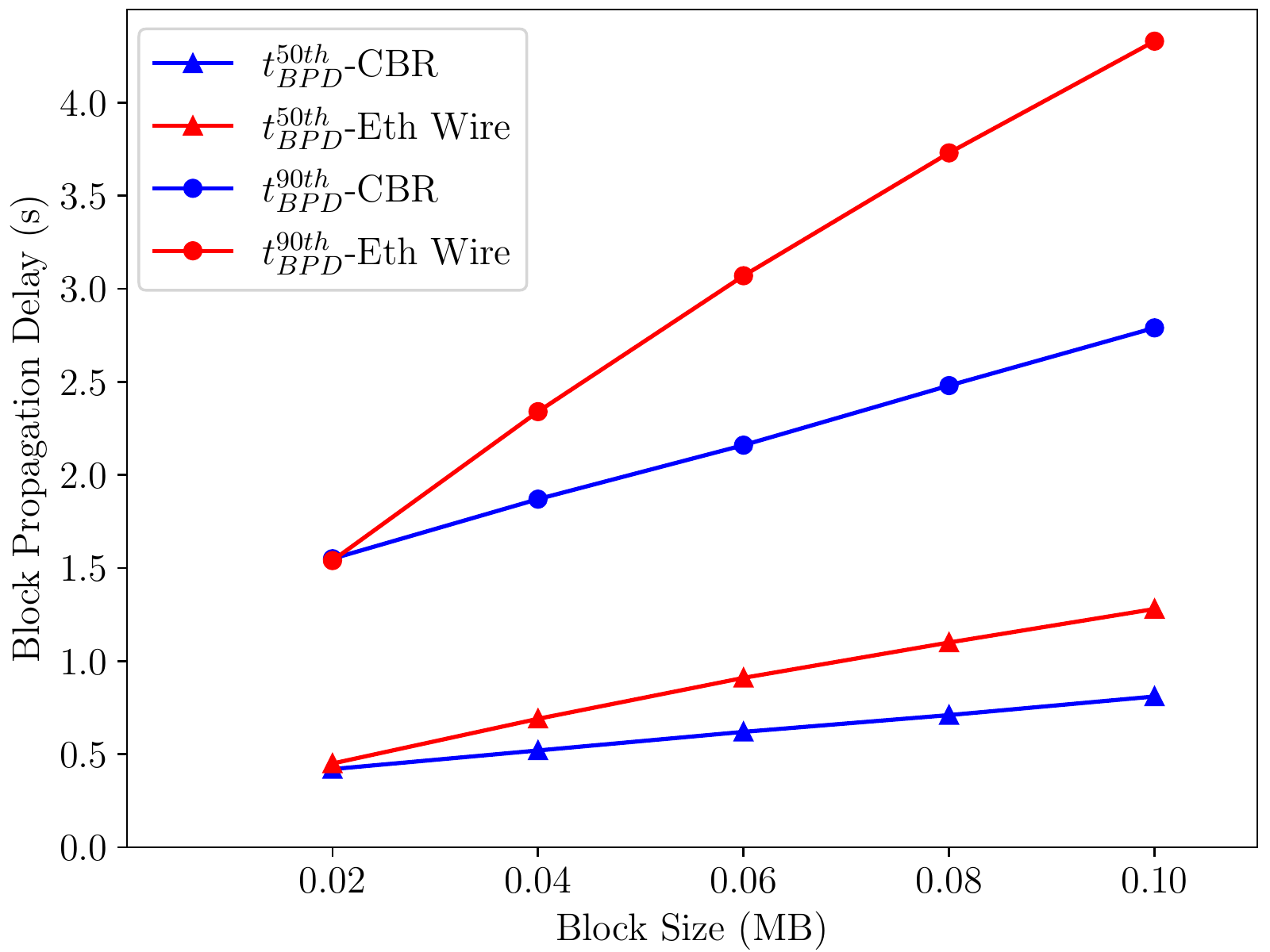}
	\caption{The 50th and 90th percentile of block propagation delay, using CBR and Eth Wire Protocol with different block sizes}
	\label{FIG:BPD}
\end{figure}

\begin{figure}[!hbt]
	\centering
		\includegraphics[width=0.8\linewidth]{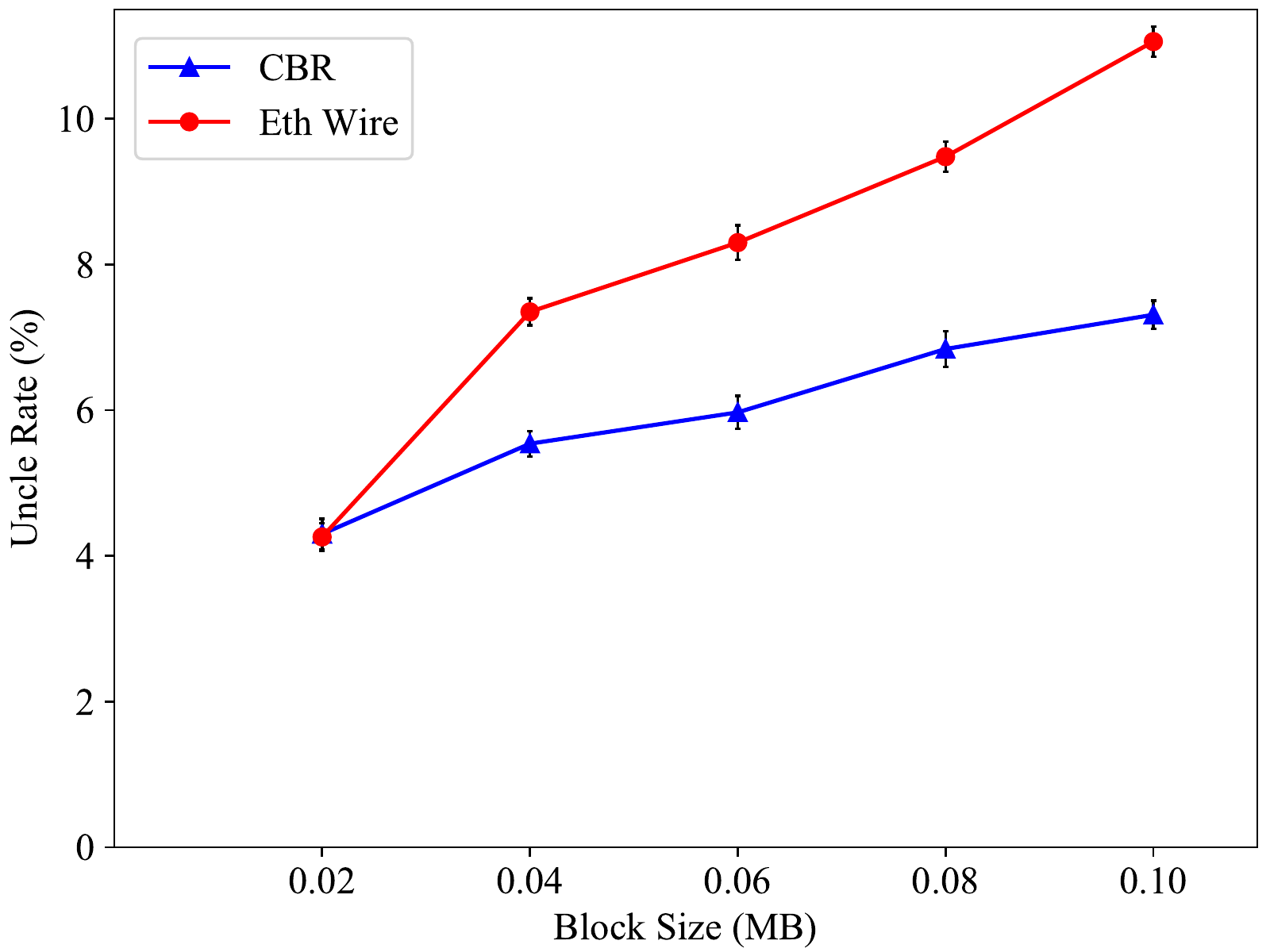}
	\caption{The uncle rate, using CBR and Eth Wire Protocol with different block sizes}
	\label{FIG:R}
\end{figure}

Bitcoin uses a compact block relay to reduce the delay caused by increased block size while Ethereum applies the {\em Ethereum wire} protocol to propagate blocks. The two protocols are illustrated in Figure~\ref{FIG:blockchainTransmission}. To compare the two protocols, we use the Ethereum configuration in Table~\ref{TB: Configuration}, vary the block size from $0.02$ MB to $0.1$ MB and load the two different protocols in the information propagation module. Figure~\ref{FIG:BPD} presents the 50th and 90th percentile of the block propagation delay. Figure~\ref{FIG:R} presents the uncle rate when using the two protocols.

When the block size is $0.02$ MB which is close to the average block size of Ethereum observed in \cite{Maeng2021} (March 2020), the two protocols have almost identical block propagation delay. As the block size increases, the block propagation delay and uncle rate grow and CBR outperforms the Ethereum Wire Protocol increasingly. Obviously, the block size has a significant impact on the block propagation delay and uncle rate. If the block size of Ethereum continues to increase, the transmission protocol should be improved in the future. For researchers or developers who aim to build an Ethereum-like blockchain, it is necessary to consider using CBR to improve the blockchain performance if the block size is rather large.

\subsection{The Impact of Average Degree}

\begin{figure}[!hbt]
	\centering
		\includegraphics[width=0.8\linewidth]{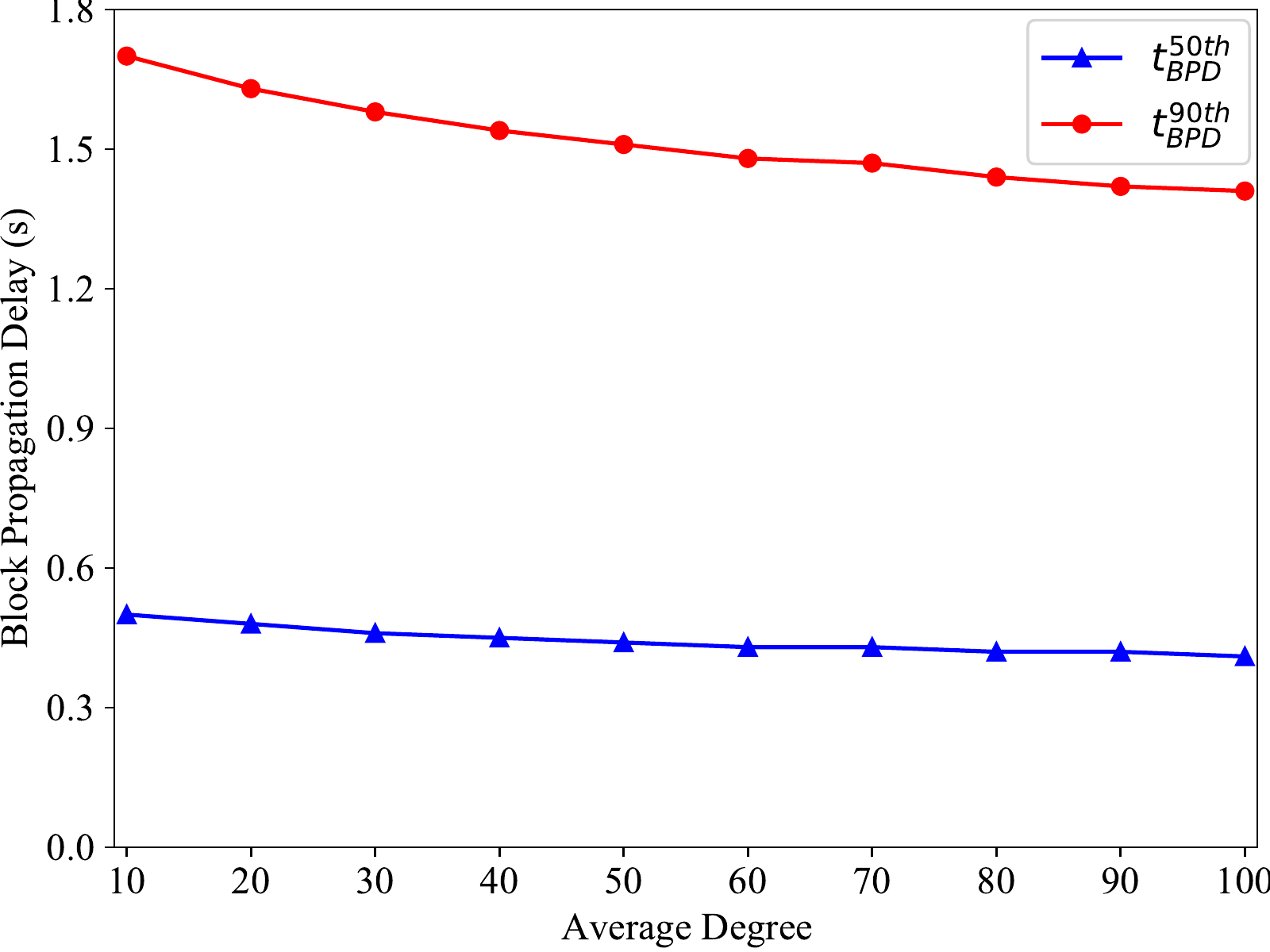}
	\caption{The impact of average degree on block propagation delay}
	\label{FIG:DegreeBPD}
\end{figure}

\begin{figure}[!hbt]
	\centering
		\includegraphics[width=0.8\linewidth]{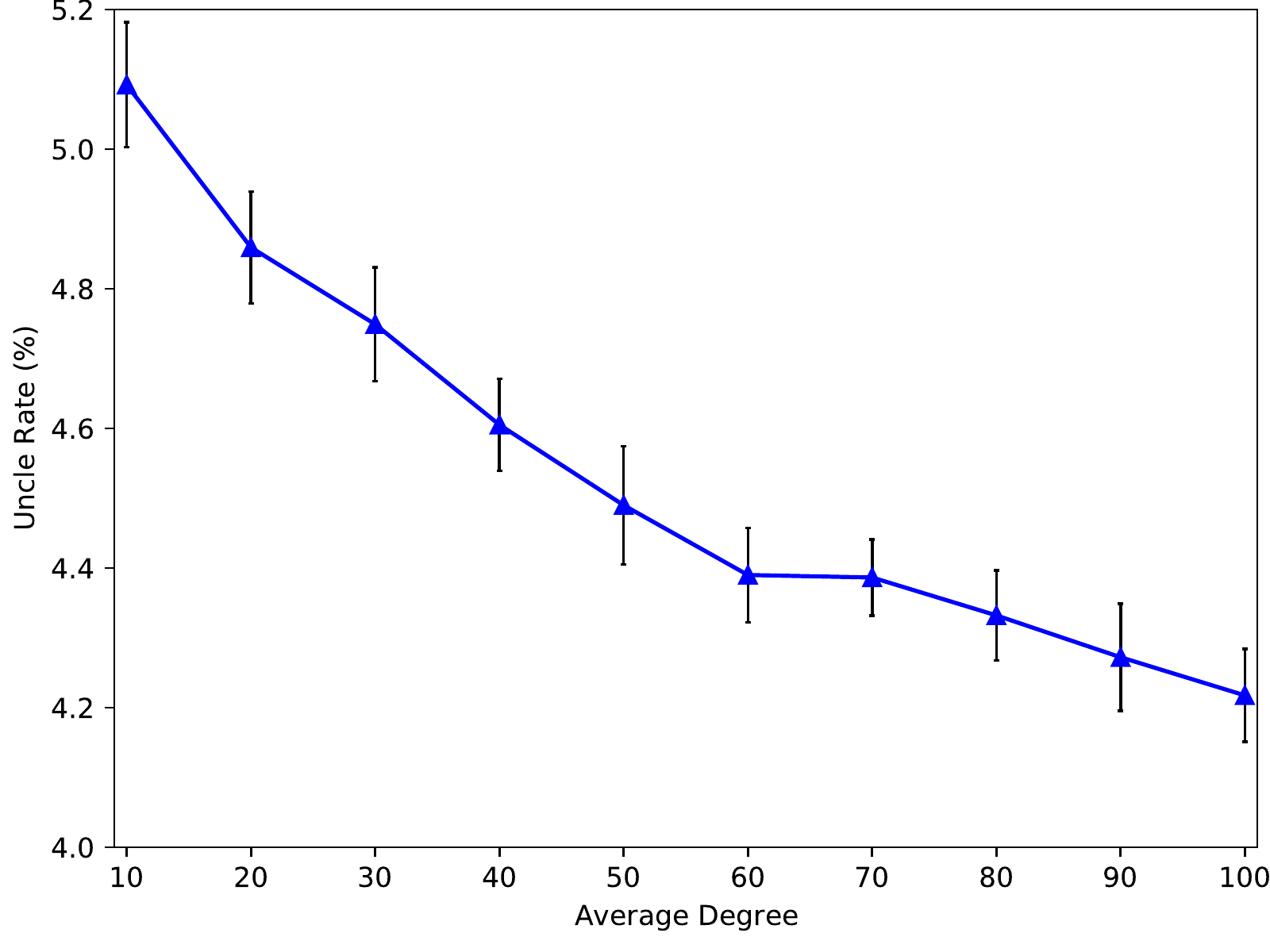}
	\caption{Impact of the average node degree on the uncle rate}
	\label{FIG:DegreeR}
\end{figure}

Besides the transmission protocols the average node degree also affects the blockchain network significantly. Increasing the average node degree reduces the block propagation delay but results in more network traffic. In this use case we use CBlockSim to investigate the impact of the average node degree on the blockchain performance. We vary the average node degree of Ethereum from $10$ to $100$ and the simulated block propagation delay and uncle rate are presented in Figures~\ref{FIG:DegreeBPD} and \ref{FIG:DegreeR}.

A continuous decrease of the block propagation delay and uncle rate can be observed as the average node degree increases. 
As the average degree increases, the uncle rate drops fast at first and then gradually slows down. When the average degree increases from 10 to 50, the uncle rate has dropped by about 0.6 percent. But when the degree increases from 50 to 100, the uncle rate has only dropped by about 0.3 percent.

\section{Conclusion}
The paper presented CBlockSim which is a modular high-performance blockchain simulator extended from BlockSim:Alharby. We publish our code on Github\footnote{https://github.com/xuyangm/CBlockSim}. Various blockchain components including different consensus protocols, network transmission protocols and finalization rules are integrated into the simulator in the form of individual modules. Users can combine existing modules flexibly or extend the simulator by adding new modules to simulate different blockchains. The design of the transaction pool structure which uses a binary sequence and bitwise operation improves the simulation performance. The simulation time is reduced by an order of magnitude and the scale of the network can be extended up to ten thousand nodes or more. With the proposed network module the simulator can be applied to network-related research which is not covered by many other simulators. Two typical network-related issues are investigated as use cases.

\end{document}